\documentclass{aa}  

\usepackage{graphicx}
\usepackage{txfonts}
\usepackage{lipsum}
\usepackage{subcaption}         
\usepackage{lscape}             
\usepackage{placeins}           
                                
\begin{document}

   \title{Massive star clusters detected by JWST as natural birth places to form intermediate-mass black holes}

   \titlerunning{Massive star clusters as birthplaces for intermediate mass black holes}


   \author{Dominik R.G. Schleicher\inst{1}
        \and Mat\'ias Liempi\inst{1} \and Mirek Giersz\inst{2} \and Marcelo C. Vergara\inst{3} \and   Francesco Flammini Dotti\inst{4,5,1} \and \newline Paulo Solar\inst{6}\and Andr\'es Escala\inst{7}\and Muhammad A. Latif\inst{8}  \and Basti\'an Reinoso\inst{9} \and Abbas Askar\inst{2} \and Raffaella Schneider \inst{1}  \and \newline Roberto Capuzzo-Dolcetta\inst{1} \and Jorge Saavedra-Bastidas\inst{10,11} \and Fernando Cuevas\inst{10} }

   \institute{Dipartimento di Fisica, Sapienza Universit\`a di Roma, Piazzale Aldo Moro 5, 00185 Rome, Italy\\  \email{dominik.schleicher@uniroma1.it}
            \and Nicolaus Copernicus Astronomical Centre, Polish Academy of Sciences, ul. Bartycka 18, 00-716
Warsaw, Poland
\and
Astronomisches Rechen-Institut, Zentrum für Astronomie, University of Heidelberg, M\"onchhofstrasse 12-14, 69120, Heidelberg, Germany
\and Department of Physics, New York University Abu Dhabi, PO Box 129188 Abu Dhabi, UAE
    \and  Center for Astrophysics and Space Science (CASS), New York University Abu Dhabi, PO Box 129188, Abu Dhabi, UAE
\and
Hamburg Observatory, Hamburg University, Gojenbergsweg 112, 21029 Hamburg, Germany
\and
        Departamento de Astronom\'ia, Universidad de Chile, Casilla 36-D, Santiago, Chile  
\and 
Physics Department, College of Science, United Arab Emirates University, PO Box 15551,Al-Ain, United Arab Emirates
\and 
     Department of Physics, Gustaf H\"allstr\"omin katu 2, FI-00014, University of Helsinki, Finland  
     \and 
     Departamento de Astronomía, Facultad Ciencias Físicas y Matemáticas, Universidad de Concepción, Av. Esteban Iturra s/n,
Concepción, Chile
\and
Universit\"at Heidelberg, Zentrum für Astronomie, Institut für Theoretische Astrophysik, Albert-Ueberle-Straße 2, 69120 Heidelberg, Germany
}

   \date{Received September 30, 20XX}

 
  \abstract
   {The James Webb Space Telescope (JWST) has detected, through gravitational lensing, several young massive star clusters (YMCs), which are considered  as relevant building blocks of high redshift galaxies. In this work, we show how a significant fraction of these YMCs could act as relevant birth places for intermediate-mass black holes. We first consider the formation of massive clusters and show that the population of YMCs is consistent with a steep mass-radius relation, which includes a relevant spread of roughly an order of magnitude. We pursue a comparison of this population with young star clusters in the local Universe and Milky Way globular clusters, including an analysis of the characteristic timescales. The YMCs show a wide spread over these properties, but include systems with both short relaxation times as well as relatively short collision timescales, implying they could go through efficient core collapse, which would lead to runaway collisions. We provide quantitative estimates of the sizes of the clusters that could efficiently form intermediate-mass black holes through a runaway collision-based channel, suggesting that these roughly correspond to the systems beyond the $1\sigma$ scatter in the mass-radius relation. This implies a fraction of $\sim16\%$ of YMCs as candidates to form intermediate-mass black holes. We show that above a mass limit of $\sim6\times10^6$~M$_\odot$, compact star clusters are likely to retain gas even in the presence of strong supernova feedback, altering the dynamics in the central core and providing the possibility to rapidly grow the central object both via gas dynamical friction and Bondi accretion. Finally, we consider the possibility of a gas-dominated regime, in which strong gravitational torques may inhibit star cluster formation and instead directly form a high-mass black holes, as suggested to have occurred in the $\infty$ galaxy.}


   \maketitle
\nolinenumbers

\section{Introduction}
The James Webb Space Telescope (JWST)\footnote{JWST: https://science.nasa.gov/mission/webb/} has produced highly revolutionary results; some were initially so surprising that it was claimed that the detected galaxies were too massive to exist at early times in the Universe \citep[e.g.][]{Labbe2023}. Over time, this issue has been better understood—for instance, it was realized that AGN contamination can make galaxies appear brighter and therefore more massive than they actually are \citep{Chworosky2024}. These investigations have shown that the galaxies are not excessively massive, but they are still more abundant than previously expected, exhibiting only a shallow evolution in volume densities. This can be explained by a star formation efficiency that increases with cloud surface density, consistent with more general theoretical expectations for the high-redshift Universe \citep{Somerville2025}.
Analyses of cloud-scale simulations from different works \citep[e.g.][]{Lancaster2021, Menon2024, Kim2018} indicate that an increase of star formation efficiency with cloud density is expected. Similarly, \citet{Haid2018} showed that the efficiency of feedback strongly decreases with increasing gas volume density.

In the early Universe, environments are likely to reach such high surface densities because the mean density of halos is generally larger, and efficient cooling at low metallicity requires higher gas densities \citep[e.g.][]{Omukai2005}. Fundamentally, both theoretical work and JWST results point to higher densities and increased star formation efficiencies at high redshift. This trend is apparent both in general statistical properties and in specific objects, such as the recently discovered Little Red Dot galaxies \citep{Matthee2024, Greene2024, Akins2025, Zhang2025}. For these objects, \citet{Guia2024} inferred very large central stellar densities by fitting a Plummer distribution to the observed stellar masses and half-mass radii, finding large central densities with a median of $\sim10^4$~M$_\odot$~pc$^{-3}$ and extending to maximum values of $\sim10^8$~M$_\odot$~pc$^{-3}$. These high densities make them ideal candidates for the formation of supermassive black holes via collision-based channels \citep{Escala2025, Pacucci2025, Dekel2025}. This channel has recently been validated via numerical simulations \citep[e.g.][]{Reinoso2018, Sakurai2019, Askar2021, Aaskar2022, Vergara2023, Vergara2025a} and was shown to provide relevant predictions for black hole masses in the local Universe \citep{Escala2021, Vergara2024, Liempi2025}. With recent work on the expected mass loss on collisions the theoretical uncertainties are being reduced and the picture becomes more firmly established \citep{Ramirez2025}.

The presence of high stellar densities very  likely is not restricted to Little Red Dots, but a more generic phenomenon in the early Universe. JWST has meanwhile detected young massive clusters (YMCs) as the building blocks of high-redshift galaxies in several environments via strong gravitational lensing. \citet{Vanzella2022a} have detected a high star cluster formation efficiency in the strongly lensed Sunburst Lyman-continuum galaxy at $z = 2.37$, with dynamical cluster masses of order $10^7$~M$_\odot$ and radii of $\sim8$~pc. At redshift $z=4$ \citet{Vanzella2022} detected massive $10^6$~M$_\odot$ clusters with radius between $3$ and $20$~pc in the Hubble Frontier Field A2744. \citet{Adamo2024} detected bound clusters in the Cosmic Gems, an ultraviolet faint galaxy at $z\sim10.2$. These clusters were shown to have radii of $\sim1$~pc and high stellar surface densities of $10^5$~M$_\odot$~pc$^{-2}$ \citep[see also][]{Messa2025, Vanzella2025}. At redshift $z=8.296$, \citet{Mowla2024} detected clusters within the Firefly Sparkle, a set of massive clusters cocooned in a diffuse arc, both with masses and radii in a very similar range. We further note the detection of very massive star forming clumps at high redshift via the gravitational lensing technique, both in the  Cosmic Grapes at redshift $z = 6.072$ \citep{Fujimoto2025} and the Misty Moons at redshift $z\sim11-12$ \citep{Nakane2025}. While these detections depend on somewhat fortunate circumstances due to the requirement of strong lensing, the fact that several such systems were found suggests a relevant role for them in the early Universe. 

The recent work by \citet{Vergara2025b} already indicated that the more compact ones among these clusters should be ideal places to form intermediate mass black holes considering stellar collisions. In addition, it is highly plausible that some of these clusters may also contain gas. The implications of black hole accretion from gas have been pointed out by \citet{Vesperini2010} and \citet{Leigh2013}, while the interplay between accretion and collisions was more systematically studied by e.g. \citet[][]{Boekholt2018, Tagawa2020, Das2021, Schleicher2022}. Simulations have shown that gas can play a strong role in the formation of very massive objects, leading to accretion-dominated scenarios as described by \citet{Chon2020, Chon2025}, or situations where accretion and collisions provide roughly comparable contributions \citep{Reinoso2023, Solar2025, Reinoso2025}. A suite of simulations has been analyzed by \citet{Saavedra2024} using machine learning techniques, showing that compactness (quantified in terms of a collision parameter) is highly favorable to form central massive objects in gas-dominated environments. Overall, it is highly plausible that such gas-based scenarios may help to explain some of the supermassive black holes detected by JWST, including some of the more extreme cases, such as the detection of an active galaxy with a black hole to bulge mass ratio of $\sim0.4$ at high redshift \citep{Juodbalis2024, Chen2025}.

In general, quite a number of potential black hole formation channels is known in the literature, including remnants from the first stars \citep{Bromm2002, Abel2002, Yoshida2008}
, the formation of massive black holes via direct collapse \citep[e.g.][]{Koushiappas2004, Bromm2003,Wise2008, Begelman2009, Schleicher2010,  Latif2013BH}, scenarios based on the collisions in dense star clusters \citep{Devecchhi2012, Reinoso2018, Sakurai2017,   Reinoso2020, Escala2021, Vergara2021,  Vergara2023, Fujii2024, Rantala2025, Rantalla2025b} as well as channels based on the contraction of black hole clusters in the centers of Nuclear Star Clusters \citep{Davies2011, Lupi2014, Kroupa2020, Chassonnery2021, Gaete2024}. We recall that even intermediate scenarios exist based on the interaction of gas with stellar dynamics \citep{Boekholt2018, Tagawa2020, Aaskar2022,Schleicher2022}.

In this paper, we provide a more systematic assessment on the potential to form intermediate-mass black holes in the dense star clusters detected by JWST. We start in section~\ref{formation} with some basic considerations concerning the formation of massive star clusters, including the implications of the physics of cooling and the expected mass-radius relation at the time of formation. In section~\ref{survival}, we compare the population of star clusters detected by JWST with different star cluster populations at the present day and with characteristic timescales for their evolution and destruction. In section~\ref{blackhole}, we provide a quantitative assessment on the potential to form intermediate-mass black holes both for the pure stellar collision channel as well as in gas-dominated scenarios. For the latter, we will propose a potential application to the active black hole detected in the $\infty$ galaxy \citep{Dokkum2025a}. Our final discussion and conclusions are provided in Section~\ref{conclusions}.

\section{Star cluster formation: cooling physics and its implications}\label{formation}

The formation of star clusters is a complex process that is highly regulated by the physics of cooling, turbulence and feedback from the process of star formation \citep[e.g.,][]{Klessen2016}. It is almost inevitable that this process will depend at least somewhat on the environment, and will be different at high redshift compared to the early Universe \citep[see e.g.][]{Clark2008, Latif2016, Latif2020, Reinoso2025}. A general model of star cluster formation so far is not available, and will likely be difficult to obtain also in the near future. Nonetheless, we can come to a certain understanding by considering what is known in the present-day Universe, and considering possible modifications in high-redshift environments. 

As a starting point, we consider the simulations by \citet{Grudic2023} as an example, who self-consistently followed star cluster formation in a Milky-Way type galaxy. The majority of their clusters were still in the range of $0.3-3$~Z$_\odot$, so we do not expect that the results were significantly influenced by chemical evolution. The mass-radius relation they obtained for their clusters is given as\begin{equation}
R_{\rm cluster}=1.4\rm{\ pc}\left( \frac{M_{\rm cluster}}{10^4\ M_\odot} \right)^{0.25},\label{massr}
   \end{equation}
with $1\sigma$ fluctuations of an order of magnitude and $2\sigma$ fluctuations of about two orders of magnitude. The relation is quite steep and suggests that the cluster radius only increases moderately with mass. However, the reported scatter is also quite significant and suggests that also very compact and dense clusters with radii of $0.1$~pc or below should occasionally form. 

In a highly complementary way, \citet{Marks2012} derived a relation to constrain the initial conditions for galactic star clusters. For this purpose, they modeled the binary evolution for 6 star clusters in Taurus to model their observed binary population, and complemented it with additional data from molecular clumps that just began forming stars \citep{Mueller2002, Shirley2003, Fontani2005}, yielding the relation\begin{equation}
R_{\rm cluster}=0.10^{+0.07}_{-0.04}\mathrm{\ pc}\left(\frac{M_{\rm cluster}}{M_\odot}  \right)^{0.13\pm0.04}.\label{Marksrelation}
\end{equation}

A very similar relation has been found by \citet{Larsen2004} for a sample of star clusters in non-interacting spiral galaxies. 

\begin{figure}[h!]
   \centering
\includegraphics[width=0.9\hsize]{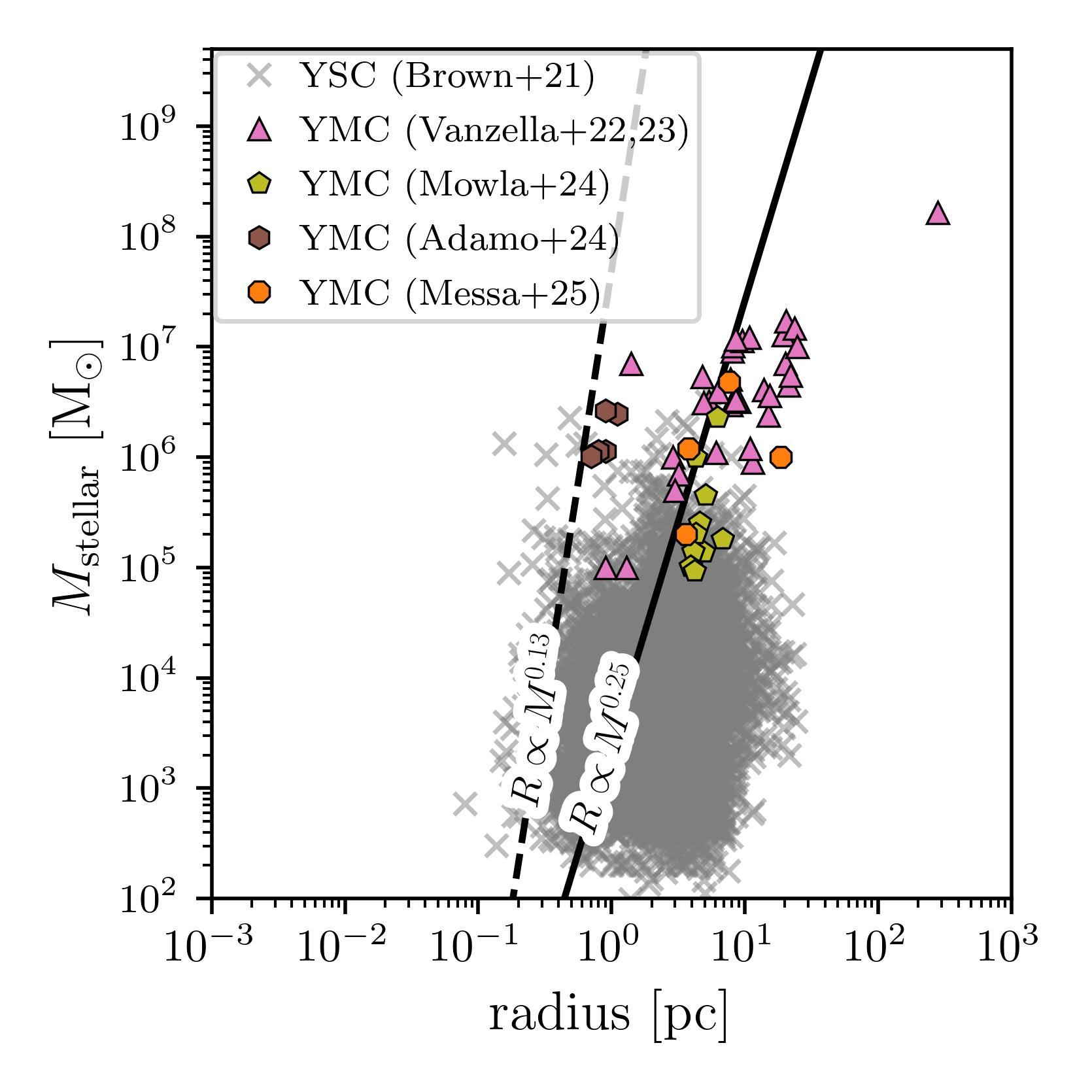}
      \caption{
YSCs in the local Universe \citep{Brown2021} and the YMCs detected by JWST in the mass-radius diagram. The \citet{Marks2012} relation (dashed line) provides a good description for the more compact clusters within the sample, while the \citet{Grudic2023} relation (solid) line provides a better description for the typical clusters. The $1\sigma$ scatter reported by \citet{Grudic2023} corresponds to an order-of-magnitude variation of the radii, implying that the observed systems are consistent with the proposed relation.}
         \label{massradius}
   \end{figure}

In Fig.~\ref{massradius}, we compare the relations from \citet{Grudic2023} and \citet{Marks2012} with the Young Star Clusters (YSCs) \citep{Brown2021} extracted from 31 galaxies from the Legacy Extragalactic UV Survey (LEGUS) \citep{Calzetti2015}, and also with the YMCs detected with JWST. The relation by \citet{Marks2012} provides a good description for the more compact clusters both among the YSCs and the YMCs, suggesting that this might indeed be the initial relation when the clusters are forming. The relation by \citet{Grudic2023} also provides a good fit but more to the average population, and the scatter within the population is also about consistent with the scattering within the \citet{Grudic2023} data, corresponding to about an order of magnitude variation in the $1\sigma$ range.

At higher redshift, the cluster formation was likely influenced by additional processes. The redshift itself can play a role, implying stronger cosmic microwave background (CMB) radiation, but also the metallicity and the dust content of the gas  \citep{Bromm2003, Omukai2005, Schneider2006}. It is also worth to recall that a continuous relation between redshift and metallicity does not exist; we have a distribution of metallicities even at redshift $0$ \citep{Tremonti2004, Berg2019, Maiolino2019} and also at high redshift, highly evolved objects like large galaxies or Active Galactic Nuclei frequently still harbor close to solar metallicities \citep{Dietrich2003, Nagao2006, Juarez2009}. At close to redshift zero, the CLASSY project\footnote{CLASSY project: (https://archive.stsci.edu/hlsp/classy} has found a variation of metallicity between $\sim5\%$~Z$_\odot$ and solar metallicity \citep{Berg2019}. Other factors could affect  star cluster formation as well, including the masses and properties of the clouds from which they form. In particular,  the star-forming clouds were likely more massive at earlier times due to the galactic environment \citep[e.g.][]{Dekel2006}.

While it is beyond the scope of the paper to analyse the resulting effects in their full complexity, we will discuss some more generic features that might be robust regardless of the circumstances. Roughly independent of metallicity and cosmic redshift, the $\tau=1$ dust opacity limit provides a relation \citep[see Eq.~20 of][]{Omukai2005},\begin{equation}
T=\left(\frac{k_B^3}{12\sigma^2m_H}  \right)^{1/5}n_H^{2/5},\label{dust}
\end{equation}
where $T$ is the gas temperature, $n_H$ the number density of the gas, $k_B$ is the Boltzmann constant, $\sigma$ the Stefan-Boltzmann constant and $m_H$ the mass of the hydrogen atom. Further assuming the presence of a temperature floor $T_{\rm min}$ before reaching the opacity limit, which may be set by the heating-cooling balance or by the CMB as a minimum temperature floor, then by equating $T_{\rm min}$ with Eq.~\ref{dust} we can derive a critical mass density  as \citep{Riaz2020}\begin{equation}
\rho_{\rm crit}=6.12\times10^{-16}\left(\frac{T_{\rm min}}{\mathrm{K}}  \right)^{5/2}~\mathrm{g\, cm}^{-3},
\end{equation}
where we employed $\rho=m_H\mu n_H$ with $\mu$ the mean molecular weight normalized by the hydrogen atom. In case that cooling is efficient and the temperature floor is indeed set by the CMB temperature, with \begin{equation}
T_{\rm CMB}=2.72(1+z)
\end{equation}
and $z$ the cosmic redshift, we find that $\rho_{\rm crit}\propto(1+z)^{5/2}$, implying a potential increase of the critical density towards higher redshift. A similar effect would also be expected in case of lower metallicity when cooling gets less efficient, as it would also tend to increase $T_{\rm min}$. Due to the significant change of the thermal evolution at $\rho_{\rm crit}$, fragmentation would be expected to occur at that density.

The density at which fragmentation occurs will also be reflected in the typical separation of fragments. If we assume such a fragment to have a Jeans mass $M_J\propto T_{\rm min}^{3/2}\rho_{\rm crit}^{-1/2}$, the number density of the fragments $n_f$ will scale as \begin{equation}n_f\propto\rho_{\rm crit}/M_J\propto \rho_{\rm crit}^{3/2}T_{\rm min}^{-3/2},\end{equation} and their typical separation as \begin{equation}\lambda_f\propto n^{-1/3}\propto \rho_{\rm crit}^{-1/2}T_{\rm min}^{1/2}.\end{equation} Again assuming that both the critical density and the minimum temperature are regulated by the CMB floor, we obtain a potential dependence $\lambda_f\propto(1+z)^{-5/4}(1+z)^{1/2}\propto (1+z)^{-3/4}$. We stress again that in reality, the temperature floor will be regulated by heating and cooling, but at least in the limiting case where cooling is still efficient, we would expect approximately this relation, while for inefficient cooling the scale should decrease faster due to the steep dependence of $\rho_{\rm crit}$ on $T_{\rm min}$. At high redshift, the physics of fragmentation would suggest clusters to be more compact compared to the present day as we discussed above.

\section{Star clusters at the present day and high redshift}\label{survival}

To assess the impact and role of the star clusters detected with JWST, we can compare their properties with these of the star cluster population in the present-day Universe and to assess the characteristic timescales within these clusters. For this purpose, we employ the sample of YSCs from the local Universe \citep{Brown2021} already introduced above; we also consider the sample of globular clusters (GCs) from \citet{Baumgardt2018} as well a sample of GCs with intermediate-mass black hole candidates compiled by \citep[][and references therein]{Vergara2024}, as well as the young massive star clusters (YMCs) detected by JWST at high redshift \citep{Vanzella2022, Vanzella2023, Mowla2024, Adamo2024, Messa2025}.

A possibly destructive effect at early stages of cluster formation is due to the gas expulsion \citep{Lada1984, Boily2003}. In massive clusters with a deep potential, this effect is however likely to be reduced. In general the implications of gas expulsion strongly depend on the dynamical state of the cluster at the time the gas is expulsed and the time scale over which it happens \citep{Goodwin2009, Smith2011, Smith2013}. Here, we focus predominantly on clusters that survive these early phases, and we consider predominantly their longer-term evolution.  The evaporation and ejection of stars is mainly caused by relaxation and tidal fields, including effects of mass segregation. This process leads to massive stars in the inner regions of the cluster, while low-mass stars are drawn in the outskirt. {For isolated clusters, the first estimates of the evaporation rate were pursued by \citet{Henon1960} and \citet{Henon1969}. More refined calculations have determined the evaporation timescale as}\begin{equation}
t_{\rm evap}=-\left( \frac{1}{N}\frac{dN}{dt}  \right)^{-1}=ft_{\rm rh},
\end{equation}
{with $t_{rh}$ the half-mass relaxation time and a numerical factor $f$. \citet{Spitzer1987} estimates $f\sim300$ considering the case of isolated cluster with single stellar mass, no binaries and disregardiing stellar evolution. For clusters influenced by a tidal field, \citet{Gnedin1999} determined $f\sim20-60$. This value might be appropriate, as many of the YMCs detected by JWST have other star clusters in their closer environment. For definiteness, we here adopt $f=40$.} We express the half-mass relaxation time as\begin{equation}
 t_{\rm rh}=\frac{0.1N}{\ln(\lambda  N)}t_{\rm cross},\label{relaxTimescale}
\end{equation}
where $t_{\rm cross}=R/\sigma$ is the crossing time determined by the half-mass radius $R$ and the velocity dispersion $\sigma$ of the average cluster In the case of a virialized star cluster, the velocity dispersion can be expressed as $\sigma^2\sim GM/R$. Note that the evaporation time scale is much longer than time scales related to relaxation processes, like mass segregation or dynamical friction, the latter expressed by \citep{Poregies2002}
\begin{equation}
    t_{\rm ms} = 3.3\frac{<m>}{m_{\rm max}} t_{\rm rh} \quad, 
\end{equation}
where $<m>$ is the cluster average stellar mass and $m_{\rm max}$ is the most massive star in the stellar population. The  factor $\langle m\rangle/m_{\rm max}$ is of the order $10^{-2}$ for a standard \citet{Kroupa2001} IMF, implying the mass segregation time to be shorter than the relaxation time of the cluster. For top heavy mass functions as considered e.g. by \citet{Hosek2019, Lu2013}, instead, the process of mass segregation is far slower.
For realistic clusters, the evaporation timescale is only an upper limit for the lifetime, as additional processes may even enhance their evolution. We can distinguish here between internal effects such as binary energy generation, dissipative processes such as dynamical friction due to the presence of gas and/or direct inelastic encounters between stars, as well as external effects due to strong tidal fields or a more global dynamical friction during the movement of the star cluster through the galaxy. 

The disruption of star clusters in the tidal field of a galaxy has been quantified e.g. by  \citet[][]{Baumgardt2003, Lamers2005}. Their models have considered star clusters consisting of stars following a \citet{Kroupa2001} initial mass function (IMF) in simulations with up to $\sim130.000$ stars. 

However, the evolution in realistic massive clusters is likely more complex. It can depend on the presence of primordial binaries, very massive stars as well as the fraction of black holes within these clusters \citep[e.g.][]{Madrid2017, Giersz2019, Wang2020, Gieles2023}. In a system consisting both of regular stars and stellar mass black holes, we need to consider the multi-component relaxation time \citep{Spitzer1971, Wang2020,  Antonini2019, Gieles2023} \begin{equation}
t_{rh,p}=\frac{t_{rh}}{\Psi},
\end{equation}
where $\Psi=\sum_k (n_km_k^2/v_k)/(\langle n\rangle \langle  m\rangle^2 \langle v\rangle )$ is a parameter that quantifies the multi-population properties, with $n_k$ being the number density of population $k$, $m_k$ its stellar mass, $v_k$ its typical velocity, and $\langle n\rangle$, $\langle m\rangle$ and $\langle v\rangle$ are the corresponding averages over all components. In the case of a standard \citet{Kroupa2001} IMF with exponent $\alpha=-2.3$ in the range between $1$~M$_\odot$ and 150~M$_\odot$, the fraction of black holes is of the order $7\%$. Both in the Arches cluster \citep{Hosek2019} and in the Galactic Center \citep{Lu2013}, the presence of a more top-heavy IMF has been suggested with $\alpha\sim-1.7$, implying a fraction of $\sim38\%$ in stellar mass black holes. It is conceivable that YMCs also have a top-heavy IMF and may harbor an enhanced amount of stellar mass black holes. Still, some part of them may also be removed via supernova kicks, so adopting a fraction of $\sim38\%$ as above will likely serve as an upper limit to evaluate possible implications.  From Fig.~3 of \citet{Wang2020}, we see that such a fraction of black holes implies values of $\Psi$ up to $5$, indicating a significant impact on the multi-component relaxation time. Here we  take into account the effect of the black hole population on the cluster dissolution time using Eq.~27 in \citet{Wang2020}, given as\begin{equation}
t_{\rm dis,bh}\sim t_{rh,bh}^xt_{\rm cross}^{1-x}\left( \frac{R_t}{R} \right)^{3/2},
\end{equation}
with $x\sim0.75$ and $R_t$ the tidal radius. This equation provides a good approximation for tidally filling clusters. For strongly tidally underfilling clusters, which in particular is the case for very dense system, this dissolution time rather represents a lower limit and the real dissolution time can be much longer. We note that in principle even more accurate expressions can be defined if more information about the different populations is available, for example with respect to the differences in their crossing times. As here we are interested in a more generic estimate, we will avoid the required additional assumptions and focus on the main dependencies. 

At the time the YMCs form, even their host galaxies will still be in the state of assembly, implying that tidal effects due to the galaxy are likely to strongly change over time. Here we consider only a simplified model, describing the tidal radius as \citep{vonHoerner1957, King1962}\begin{equation}
R_t=R_G\left( \frac{M_{\rm cluster}}{3M_G(R_G)}\right)^{1/3}=R_G\left( \frac{GM_{\rm cluster}}{3R_GV_C^2(R_G)} \right)^{1/3},
\end{equation}
where $M_G(R_G)$ is the galactic mass enclosed in radius $R_G$ and in the second equality we have employed the circular velocity $V_C^2(R_G)=GM_G(R_G)/R_G$. Within this framework, we can jointly assess the implications of the ratio between the cluster half-mass radius and the tidal radius, as well as the implications of a potentially enhanced population of black holes within massive clusters.

Systems for which the collision time is equal to - or shorter - than the age of the Universe are expected to form a massive central object \citep[see e.g.][]{Chassonnery2021,  Escala2021,Vergara2023, Vergara2024}. We assume here that the systems are virialized, allowing us to express the global collision time as \begin{equation}
t_{\rm coll}=\sqrt{\frac{R}{GM(n\Sigma_0)^2}},\label{colltime}
\end{equation}
where $n$ is the number density in the star cluster and $\Sigma_0$ the effective cross section due to gravitational focusing. Assuming the mean stellar mass in the cluster to be equal to a solar mass, we have an average number density $n=\frac{3M}{M_\odot 4\pi R^3}$ and $\Sigma_0=16\sqrt{\pi}R_*^2(1+\Theta)$, with $R_*$ the mean stellar radius and $\Theta=9.54\left(\frac{R_*M_\odot}{R_\odot M_*}\right)^{-1}\left( \frac{100\mathrm{km/s}}{\sigma}\right)^2$. While we have seen in previous studies that the global collision timescale  can be a good indicator for the potential of a system to form central massive objects \citep[e.g.][]{Escala2021, Vergara2023, Vergara2024}, it is also instructive to consider the timescales in the central core of the cluster. In the absence of internal energy sources, the ratio between the core radius $r_c$ and the half-mass radius $R$ can be very small \citep{Cohn1980}. In real star clusters, binaries are expected to form during the gravitational contraction, and the size of this core is generally expected to be regulated by the dynamical energy balance between the heat generated in the core (e.g. hard binaries) versus the conductive energy outflow from the cluster. This balance typically implies a ratio of core radius $r_c$ to half-mass radius $R$ in the range of $0.05-0.3$ \citep{Henon1961, Henon1965, Heggie2003, Gieles2011, Alexander2014b}. In the following, we parametrize the ratio $r_c/R=\lambda$, and adopt the density profile of the non-singular isothermal sphere. This profile does not have an exact analytic solution, but we can approximately include a central core and account for the correct asymptotic scaling with the expression\begin{equation}
\rho(r)=\rho_0\left( 1+\left(\frac{r}{r_c}  \right)^2 \right)^{-1}.
\end{equation}
We note that the difference between this expression and the exact solution is less than $10\%$ for up to 4.5 the core radius. It increases at larger radii but still allows us to roughly estimate the central density in the system considered here.  We  assume a truncation radius equal to twice the half-mass radius $R$. This implies a central density\begin{equation}
\rho_0=\frac{M_{\rm cluster}}{4\pi \lambda ^3R^3(2/\lambda-\arctan(2/\lambda))},
\end{equation}
which we convert into a number density $n_0$ assuming a mean mass per star $M_*\sim M_\odot$. The velocity dispersion in the isothermal sphere on the other hand is constant and we can evaluate it just considering global properties as\begin{equation}
 0.5\sigma^2 = 0.4\frac{GM}{R}.
\end{equation}

The cross section $\Sigma_0$ should be evaluated as indicated above. Using these expressions for central number density $n_0$, velocity dispersion $\sigma_c=\sigma$ and cross section $\Sigma_0(\sigma_c)$, the collision time in the core follows from the mean-free-path approximation as\begin{equation}
t_{\rm coll, core}=\frac{1}{n_c\Sigma_0(\sigma_c)\sigma_c}.\label{collcore}
\end{equation}

\begin{figure}[h]
    \centering
    \includegraphics[scale=0.76]{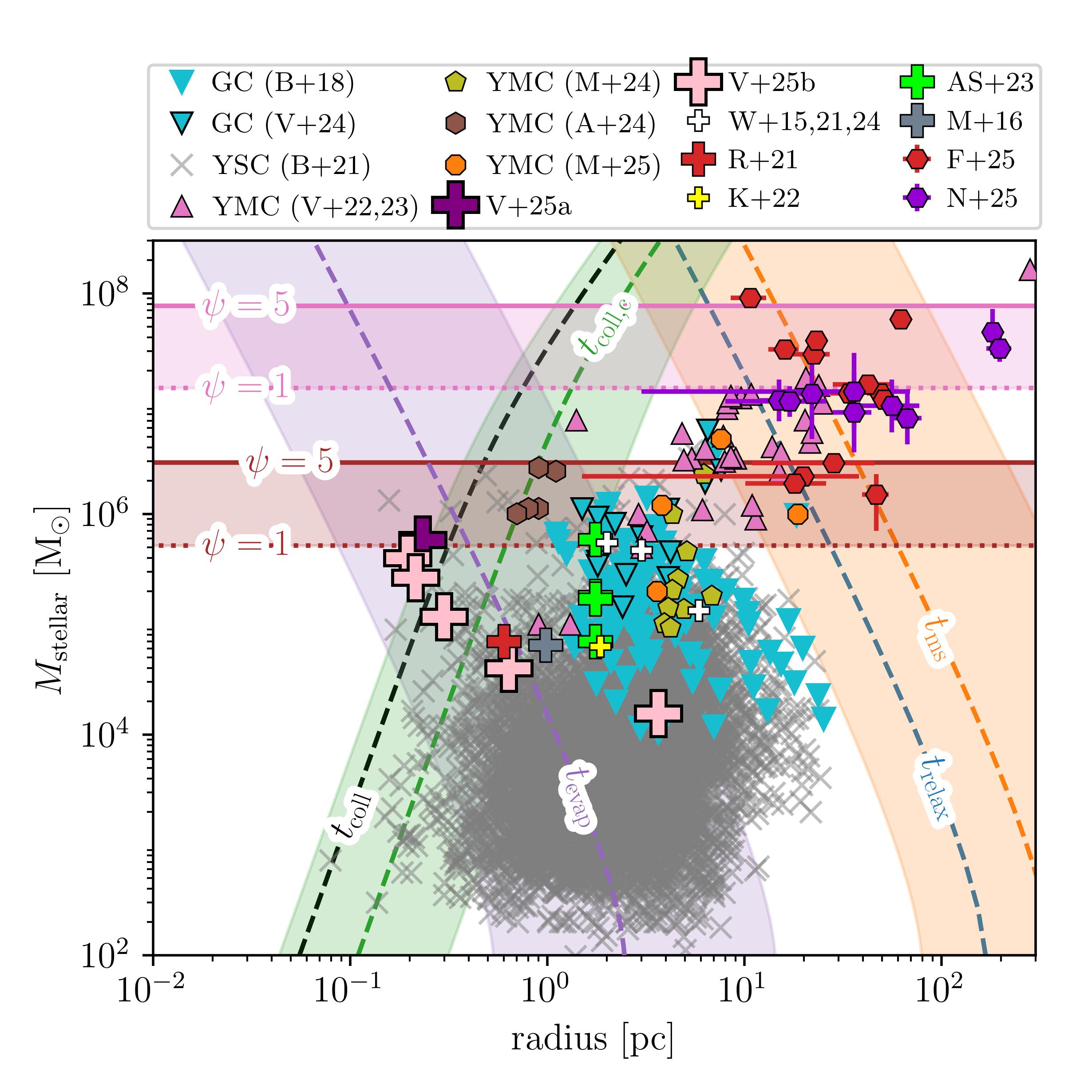}
    \caption{Populations of YSCs \citep{Brown2021}, GCs \citep{Baumgardt2018, Vergara2024}, YMCs \citep{Vanzella2022, Vanzella2023, Adamo2024, Mowla2024} and star-forminig clumps in high-redshift galaxies \citep{Fujimoto2025, Nakane2025}    in the mass vs half-mass-radius diagram. We show them together with several characteristic timescales which are evaluated for a typical value of $1.0$~Gyr (dashed lines) and minimum and maximum values of $0.1$~Gyr and $14$~Gyr, respectively. This includes the evaporation timescale (shaded purple), the dynamical friction inspiral timescale (shaded orange) and the collision timescale in the inner core (shaded green). {We further provide the relaxation time at $14$~Gyr (blue dashed line) and the global collision time at $1.0$~Gyr (black dashed line).} The minimum cluster mass to survive tidal disruption for $14$~Gyr is indicated via the red shaded area for a galactic radius of $10$~kpc,  and for a galactic radius of $1$~kpc via the pink shaded area, {assuming a circular velocity of $V_G=200$~km/s and $\Psi$ values between $1$ and $5$. Star cluster simulations are indicated as crosses, with the size of the cross indicating the mass of the massive black hole formed in the simulations  of \citet{Vergara2025a,Vergara2025b,Mapelli2016, Rizzuto2021, Arca2023, Wang2015, Wang2021, Kamlah2022, Wang2024}.}}
    \label{200}
\end{figure}

We compare here the previously introduced star cluster populations both with each other and with the characteristic timescales introduced above, as well as the star-forming clouds described by \citet{Fujimoto2025} and \citet{Nakane2025}. In Fig.~\ref{200}, the masses of these clusters are given as a function of their radius, and it is clearly visible that the YMCs correspond to the high-mass range of GCs and beyond, extending up to $\sim10^7$~M$_\odot$. 

The shaded purple region indicates the parameter space corresponding to evaporation times of the cluster between $14.0$~Gyrs and $0.1$~Gyrs. {We find that some of the more compact clusters are potentially subject to evaporation. From the YSCs, some extremely compact ones with radii of $\sim0.2$~pc and masses of $\sim10^2-10^4$~M$_\odot$ have very short evaporation timescales of less than $0.1$~Gyr. A relatively substantial fraction of about half of the YSCs have evaporation timescales of $0.1-1$~Gyr, while the remaining ones have evaporation times up to $14$~Gyr or even larger. In the case of GCs, about half of them have evaporation timescales of $1-14$~Gyr (particularly for radii of $\sim1-3$~pc), while many of them also have evaporation times longer than the age of the Universe. In case of the JWST clusters, only the most compact ones with radii of less than a parsec \citep{Adamo2024} have evaporation timescales shorter than the age of the Universe, while almost all of them have larger evaporation timescales.}

The evaporation however is only relevant for isolated clusters. Additional constraints can be obtained from tidal dissolution. The red shaded area indicates the minimum masses required to survive tidal disruption for $14$~Gyrs at a galactic radius of $10$~kpc, assuming a circular velocity of $200$~km/s and values of $\Psi$ between $1$ and $5$, while the pink shaded area indicates the same at a galactic radius of $1$~kpc. At $10$~kpc, the lower mass limit is $\sim4\times10^5$~M$_\odot$ for $\Psi=1$ (small fraction of black holes, standard Kroupa IMF) and $\sim3\times10^6$~M$_\odot$ in case of $\Psi=5$ (top-heavy IMF). Even for $\Psi=1$, the corresponding line is in the upper range of the YSCs, so that most of these would be destroyed over a timescale of $14$~Gyrs, and also a relevant fraction of GCs would be severely affected. On the other hand, most of the YMCs detected by JWST are above that line in case of $\Psi=1$, though they would be affected in case of a more top-heavy IMF with $\Psi=5$.

Within Fig.~\ref{200}, we further show the relaxation time at $1$~Gyr as the blue line. We find that most but not all of the YMCs have relaxation times less than $1$~Gyr, and certainly all of the YSCs and GCs have shorter relaxation times. Particularly some of the YMCs at very high masses and larger radii however also need very long timescales for relaxation. We compare with the parameter space for a dynamical friction mass inspiral timescale of $0.1-14$~Gyrs (shaded orange). All of the GCs and YSCs and also some of the more compact YMCs have mass inspiral times of less than $0.1$~Gyr, though for some of the more extended and massive ones these can increase up to roughly a Gyr. 

The global collision timescale at $1$~Gyr is indicated as a black dashed line, suggesting that only few of the YSCs have a global collision timescales of less than $1$~Gyr. The parameter space for a  core collision time of $0.1-14$~Gyrs is shown as the green shaded area assuming $\lambda=0.05$. It overlaps with some of the more compact YSCs, GCs and YMCs. While these collision timescales technically are still high, the fact that these clusters have short relaxation times and can undergo core collapse will significantly enhance the potential for collisions.

When we consider the star-forming clumps in the Cosmic Grapes \citep{Fujimoto2025} and Misty Moones \citep{Nakane2025}, we note that these clouds are in the mass range of $10^6-10^8$~M$_\odot$ though with larger radii, typically in the range of $10$~pc up to $100$~pc. Assuming they will maintain similar properties also after forming a star cluster, it implies that they will have longer relaxation times comparable to or larger than the age of the Universe and also larger collision times. We note that these clouds tend to extend the parameter space covered by the JWST clusters towards somewhat larger masses and larger radii. Now assuming a star formation efficiency around $10\%$ and stars to preferentially form in the centers of these clouds, it is at least possible that these populations are compatible with each other and perhaps the clouds will form dense stellar systems comparable to the YMCs. 

For comparison, we also indicate several direct N-body simulations of large star clusters from the literature, with the size of the symbol reflecting the mass of the central massive black hole that formed. In particular, the simulations by \citet{Vergara2025a, Vergara2025b} have found the formation of massive black holes with a few times $10^4$~M$_\odot$ for star clusters with masses of $10^5-10^6$~M$_\odot$ and final cluster radii of  $\sim0.2-0.7$~pc, overlapping with some of the more compact YSCs. The most compact YMCs reported by \citet{Adamo2024} have similar radii but larger masses, and may be expected to form even more massive black holes. In the simulations by \citet{Mapelli2016, Rizzuto2021} and \citet{Arca2023}, intermediate-mass black holes of a few times $100$~M$_\odot$ are formed, in clusters typically of $10^5-10^6$~M$_\odot$ with radii of $\sim1$~pc. \citet{Wang2015,Wang2021,Kamlah2022, Wang2024} report clusters that are more extended with typical sizes of a few parsec in a similar mass range, where intermediate-mass black holes do not form. From this tentative comparison, we can roughly infer the compactness that is required to form massive black holes, which is reached for the more compact clusters close to the \citet{Marks2012} relation. As the detection of YMCs via JWST is difficult and restricted to cases of strong gravitational lensing, we naturally do not have as much statistics for these clusters as for the YSCs in nearby galaxies. Nonetheless it is at least highly plausible that a similar variety of masses and radii should be expected.

\section{Massive black hole formation in dense massive clusters}\label{blackhole}

In the following subsections, we consider the formation of intermediate-mass black holes in the dense stellar clusters detected by JWST considering the formation channel based on stellar collisions, the possibility of black hole formation in case of the retention of gas in very massive systems as well as in situations that are strongly dominated by gas.

\subsection{Massive black hole formation in purely stellar systems}

\begin{figure}[h!]
    \centering
    \includegraphics[width=0.85\hsize]{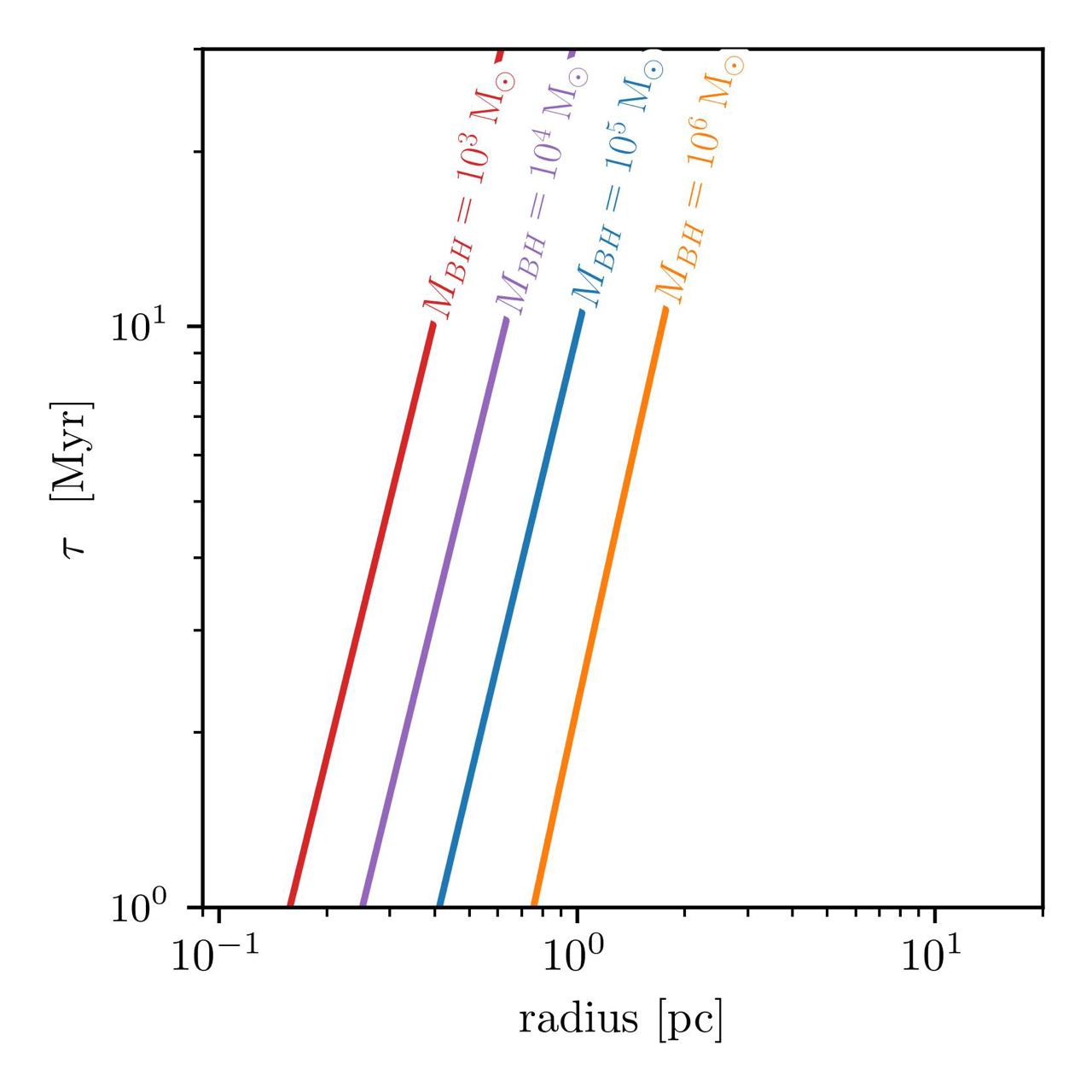}
    \caption{Assuming collision-based black hole formation, we provide the required evolutionary timescale of stellar clusters as a function of their radius to form a seed black hole with a mass as indicated by the different lines, considering seed masses from $10^3$~M$_\odot$ up to $10^6$~M$_\odot$. The seed mass is translated into a cluster mass via Eq.~\ref{seed}, and Eq.~\ref{crit} is then employed to derive the relation between cluster radius and evolutionary time, assuming that the average mass per star corresponds to a solar mass and the average radius per star to a solar radius.}
\label{fig:timerad}
\end{figure}

In purely stellar systems (no gas), massive objects can form in case of run-away collisions of stars \citep[e.g.][]{Poregies2002}. For the formation of central massive objects, \citet{Escala2021} pointed out that supermassive black hole formation in NSCs appears to be regulated by the collision timescale given in Eq.~\ref{colltime}, and also large suits of Monte-Carlo simulations have demonstrated the feasibility to form massive black holes in NSCs \citep{Giersz2015, Hong2020,Askar2021, Aaskar2022}. Particularly, NSCs without a central massive black hole all have collision times longer than their age, implying stability versus collisions, while systems in which a massive black hole has formed have collision timescales shorter than the age of the system, implying a global instability within the stellar system. Based on this consideration, a critical mass  was defined by \citet{Vergara2023}, for which the collision time defined in Eqs.~\ref{colltime}, \ref{collcore} is equal to the age of the system. It is given as\begin{equation}
M_{\rm crit}=R^{7/3}\left( \frac{4\pi m_*}{3\Sigma_0 \tau G^{1/2}}  \right)^{2/3},\label{crit}
\end{equation}
where $m_*$ is the mean mass per star in the system. Through comparison with a large suite of numerical simulations and stellar systems in the local Universe \citep[e.g.][]{Poregies2002,Arca2023, Arca2024, Mapelli2016}, it was shown that there is a relation between the black hole formation efficiency $\epsilon_{BH}$, which is defined as the ratio between the black hole mass and the final total mass of the cluster and black hole mass \citep{Escala2021, Vergara2023} and the ratio $M/M_{\rm crit}$. {We emphasize that this relation has also been validated by \citet{Vergara2023, Vergara2024} through the comparison with observed stellar systems such as Nuclear Star Clusters, with typical ages of the order of Gyrs. This suggests  that stellar evolution effects should not strongly alter this relation, even if so detailed stellar dynamical simulations for this regime are still missing. A fit to this relation has been provided by }\citet{Vergara2025b} and is given as\begin{equation}
\epsilon_{BH}\left(\frac{M}{M_{\rm crit}}\right)=\left[1+\mathrm{exp}\left( -4.63\left(\ln\left( \frac{M}{M_{\rm crit}} \right)-4  \right)  \right)  \right]^{-0.1}.
\end{equation}
When $M/M_{\rm crit}\sim0.38$, we have $\epsilon_{BH}\sim0.1$. This corresponds already to a rather unstable situation, and for more realistic clusters, we consider a more moderate efficiency $\epsilon_{BH}\sim2\%$, { which is reached for $M/M_{\rm crit}\sim0.012$}. This corresponds to the following condition to form black hole seeds at this efficiency level:\begin{equation}
\frac{M_{BH}}{0.02}=M=0.012\,M_{\rm crit}.\label{seed}
\end{equation}
If $M_{\rm crit}$ is set via Eq.~\ref{seed}, Eq.~\ref{crit} implies a relation between the radius of the star cluster $R$ and the evolutionary timescale $\tau$ needed to form a central massive object via collisions, assuming fixed values of $\Sigma_0$ and $m_*$. For simplicity we assume here that on average stars in the cluster have a solar mass and a solar radius. For this case, the resulting relation between cluster radii and the necessary evolution timescales is given in Fig.~\ref{fig:timerad} for different black hole seed masses, while the cluster mass (following Eq.~\ref{seed}) corresponds to the mass of the seed black hole. {For black hole seeds of  $\sim10^5$~M$_\odot$, the formation is feasible for cluster radii of  $\sim1$~pc within $\sim10$~Myr.} { It is worth noting that \citet{Vergara2024} also found a relevant scatter in the efficiency corresponding to at least a factor of a few, translating into a similar scatter both within the timescales and radii considered here.} In case of larger seed masses of $10^6$~M$_\odot$, the same timescale can be achieved at somewhat larger radii as the clusters are getting more massive, implying a decrease in the crossing time. Thus, the formation of $\sim10^5$~M$_\odot$ black holes will be feasible in clusters of a few parsec size and with stellar masses of $\sim3\times10^7$~M$_\odot$. On timescales of a few ten million years,   most stars above $10$~M$_\odot$ will evolve into a black hole and lose a relevant fraction of their mass, implying at least some moderate mass loss and resulting expansion within the cluster {on  timescales comparable to the formation of the black hole seed.  So far, the interaction of such stellar mass black holes with the rest of the cluster has not been explored in detail, and thus still forms a relevant uncertainty. Also the possible result of mergers of stellar mass black holes with a very massive star and its implications for the subsequent evolution needs to be further explored. We therefore limit the extrapolation here to  $\sim30$~Myrs.}  

Comparing these estimates with the mass-radius relation in Fig.~\ref{massradius} and considering that \citet{Grudic2023} derived a $1\sigma$ scatter corresponding to one order of magnitude in radii, we realize that these estimates correspond roughly to the boundary between the $1\sigma$ and $2\sigma$ range. This is further consistent with the comparison between observed and simulated clusters towards the end of section 3, which also indicated that the more compact ones are those that are well-suited to form intermediate-mass black holes. As here we only consider the deviations towards smaller radii, this implies that about $16\%$ of real star clusters should lie in the parameter space where the formation of a central massive object is possible. We checked that this ratio is also roughly compatible with the number of YMCs detected by JWST for which we expect that massive black holes could form (requiring a relaxation time of about $400$~Myr.

\subsection{Gas retention in compact massive systems and further implications}

We have so far considered the limiting case of a purely stellar systems. Detailed simulations of star clusters formation in cosmological simulations have been pursued by different groups including supernova and pre-supernova feedback \citep{Brown2022, Polak2024, Calura2025}. Typically these consider star cluster masses up to a few times $10^6$~M$_\odot$, covering essentially the lower-mass range of the YMCs discovered by JWST. Given the high masses and compactness of the YMCs, as well as the general steep relation by \citet{Marks2012} for the initial stage of the clusters, it is conceivable that not all gas will be able to escape and one may rather have to consider the possibility of having at least some retention gas within the inner part of the system. 

{We focus here predominantly on systems with metallicities  $Z\lesssim 0.1$~Z$_\odot$, as typical for the observed JWST clusters, where we also expect that stellar mass loss due to feedback will be less relevant as compared to high metallicity cases \citep[see e.g.][]{Vink2001, Vink2005, Mokiem2007, Trani2014, Das2021a, Spera2017, Spera2019}. } We  adopt here the configuration with $M=10^6$~M$_\odot$ and $R=0.6$~pc as a somewhat idealized case to explore some possible implications of such compact and massive systems.  Such a system has an escape velocity $v_{\rm esc}=\sqrt{2GM/R}\sim120$~km/s and a binding energy of $E_b\sim\frac{3}{5}GM^2/R\sim 8.6\times10^{52}$~erg. Such clusters may thus retain the winds from AGB and red-supergiant stars of $\sim10-30$~km/s, while solar-type winds with $400-800$~km/s or winds of massive stars would be expected to escape from the cluster. Within the core of the cluster, the escape velocity will be enhanced further potentially allowing to keep some of the material from the winds of the very massive stars.

In case of a standard \citet{Kroupa2001} IMF, we would expect one type II supernova explosion per $100$~M$_\odot$ in stars, thus implying at least $10^4$ type II supernovae in a cluster with $10^6$~M$_\odot$. This could be easily enhanced by a factor of $4-5$ in case of a top-heavy IMF. With an energy per supernova of $E_{SN}\sim10^{51}$~erg and a coupling efficiency $\epsilon_{SN}$, we would need a rather low coupling efficiency of $\epsilon_{SN}\sim0.01$ in order to retain gas after the supernova explosions. We note that the star formation efficiency in these environments may reach up to $60\%$ per free-fall time \citep{Polak2024}, depleting a relevant part of the gas via star formation unless it is replenished by inflows. It is  instructive to assess whether a star cluster can become robust against supernova feedback, as the retention of gas could then substantially alter the dynamics in their interior. We consider for this purpose a cluster following the \citet{Marks2012} relation, and adopt here the limiting case where the binding energy of the cluster should be equal to the effective energy provided by supernova feedback:\begin{equation}
\frac{3}{5}\frac{GM^2}{0.1\mathrm{\ pc}(M/M_\odot)^{0.13}}=\epsilon_{SN}\frac{M}{100\,M_\odot}E_{SN}.
\end{equation}
Solving for the mass $M$, we obtain\begin{equation}
M=6\times10^6M_\odot\left( \frac{\epsilon_{SN}}{0.1} \right)^{1.15},
\end{equation}
suggesting that supernova feedback may become ineffective around a mass scale of $\sim6\times10^6$~M$_\odot$. This scale of course is not completely fixed, but depends both on the feedback efficiency and could also increase in case of a more top-heavy IMF. 

It is instructive  to consider the implications of a system above that mass limit, where gas is retained even in the presence of strong feedback. For such a system we assume that in addition to the stellar mass $M$, a gas mass $M_g=0.5M_*$ is also still present. We model the density profile of the gas and the stars via a non-singular isothermal sphere with a core radius $r_c=\lambda R$ and a truncation radius of $2R$ as before in section 3, with $\lambda\sim0.1$. For $\lambda=0.1$, we obtain a stellar mass in the core of $M_{*,c}=7.0\times10^4$~M$_\odot$ and a gas mass in the core of $M_{g,c}=3.5\times10^4$~M$_\odot$.  An immediate consequence of the gas is the presence of dynamical friction, which we evaluate as \citep{Chandrasekhar1943, Ostriker1999, Binney2009}\begin{equation}
t_{\rm df,gas}=\frac{\sigma^3}{4\pi G^2m\rho\ln\Lambda},\label{dfgas}
\end{equation}
where $\sigma$ is the typical stellar velocity that we derive from virial equilibrium, $m$ the mass of the star under consideration, $\rho$ the mass density of the gas and $\ln\Lambda\sim10$ the Coulomb logarithm.  For stars with $100$~M$_\odot$, we obtain a gas dynamical friction timescale of $1.6\times10^ 6$~yr, and even for $10$~M$_\odot$ stars, it is still $1.6\times10^ 7$~yr, implying that the stars would migrate to the center within their expected lifetime. {We note that a more detailed treatment of gas dynamical friction has been provided by \citet{Ostriker1999} distinguishing between subsonic and supersonic flows, which we do not adopt here as it would require additional assumptions in our treatment.} In case of increasingly massive but still compact clusters, we note that the virial velocity may increase considerably with a corresponding significant increase in the gas dynamical friction timescale. In that case, while the dynamical friction would become less important, the gas retention itself could even become more relevant and particularly in the central region, if the gas collapses gravitationally, even additional effects such as gravitational torques may become relevant, which we discuss in more detail in the follow subsection. 

While in general any interaction between gas and stars will lead to kinetic energy dissipation and faster cluster collapse, it is useful to consider  the effects of additional processes: Stars within the cluster may  grow via Bondi-Hoyle-Lyttleton accretion with an accretion rate \citep{Bondi1944, Bondi1952}
\begin{equation}
\dot{m}_B=\frac{4\pi G^2m^2\rho}{(v_{\rm rel}^2+c_s^2)^{3/2}},
\end{equation}
where $v_{\rm rel}$ is the relative velocity between gas and stars (estimated here based on virial equilibrium), adopting a generic sound speed $c_s=10$~km/s appropriate for photo-ionized gas. For a star with $100$~M$_\odot$, it implies a timescale for Bondi accretion  of $m/\dot{m}_B=1.7\times10^7$~yrs. This timescale is inversely proportional to the $m^2$, so if even a few massive stars merge in the center of the core, the Bondi accretion process could become highly efficient and 
a large part of the total gas mass in the cluster could be accreted onto the central object within a short time \citep[see also][]{Vesperini2010, Leigh2013, Roupas2019}.

Another potential competing process could be star formation. In the central core we have a gas surface density of \begin{equation}
\Sigma_g=M_{g,c}/(\pi \lambda^2 R^2)=1.9\times10^6 M_\odot \mathrm{pc}^{-2},\end{equation} 
where we calculated the half-mass radius $R$ from the \citet{Marks2012} relation. Applying a Kennicutt-Schmidt law with \begin{equation}\Sigma_{\rm SFR}[M_\odot\,\mathrm{yr}^{-1}\,\mathrm{pc}^{-2}]=2.5\times10^{-4}\left( \Sigma_g [M_\odot\,\mathrm{pc}^{-2}] \right)^{1.4},\end{equation} \citep{Kennicutt1998, Bigiel2008} we obtain a gas depletion timescale  $\Sigma_g/\Sigma_{\rm SFR}\sim1.2\times10^7$~yr. While this suggests that initially the star formation timescale could be comparable to the timescale for accretion, at least once that a central massive object is in place, it is very likely that accretion would become dominant as a runaway process. The timescale for Bondi accretion is very similar to the gas dynamical friction timescale for $10$~M$_\odot$ stars. If the gas is not depleted but there remains a continous inflow from the outside, then within $1.6\times10^8$~years, even the solar mass stars in the core would  migrate to the center and merge with the central object. Under the somewhat conservative assumption of a \citet{Kroupa2001} IMF, $\sim73\%$ of the stellar mass in the core would be expected to merge with the central object, while in addition it is very likely that the gas mass in the core and potentially even from the outer parts of the cluster could be accreted. It is then conceivable that a supermassive black holes of a few times $\sim10^5 - 10^6$~M$_\odot$ or more could form within such a cluster, depending on specific properties and if/inhowfar accretion onto the cluster is still ongoing.

{For high metallicity systems ($Z\gtrsim0.1$~Z$_\odot$), some differences can potentially be expected. Feedback from winds is generally stronger at high metallicity, as verified for winds from O and B stars by \citet{Kudritzki2000,   Vink2001, Mokiem2007, Puls2008, Tramper2011}  for Wolf-Rayet winds by \citet{Vink2005}, for winds from Wolf-Rayet nitrogen-type (WN) stars by \citet{Graefener2008} and for Red Supergiants and AGB stars by \citet{vanLoon2005}. Due to such processes, even the masses of stellar mass black holes appear to be higher in lower-metallicity environments \citep{Belczynski2010}. At high metallicity, feedback from winds may thus contribute more strongly to gas expulsion \citep{Das2021a}, and the stellar mass loss through winds could act as an energy source and partially substitute binary hardening as a heating mechanism \citep{Trani2014, Spera2017, Spera2019}.  Here we verified using an updated version of MOCCA with the single and binary stellar evolution synthesis tracks from \citet{Tanikawa2020} that mass loss occurs more rapidly around solar metallicity reaching $\sim10\%$ contributions within less than $10$~millions years, and fractional mass losses of $\sim30\%$ over timescales of $200$~million years. The resulting implications thus need to be taken into account in the dynamics. Nonetheless, \citet{Rantala2026} demonstrated that the formation of very massive stars is still feasible even at solar metallicity, even if the masses of the resulting objects are likely reduced by a factor $4-5$. }

\subsection{Massive black hole formation in strongly gas-dominated  clusters and application to the $\infty$ galaxy}

The scenario sketched in the previous subsection could become more extreme in a case where the system is more strongly gas-dominated. While in subsection~4.2 we assumed the gas mass to correspond to about $50\%$ of the stellar mass, it is conceivable that more extreme scenarios will occur, for example in the context of galaxy mergers \citep[e.g.][]{Mayer2015, Mayer2024}. A possibly intriguing site for such a scenario has recently been identified by \citet{Dokkum2025a, Dokkum2025b} in the $\infty$ galaxy, a $z=1.14$ object imaged by JWST in the COSMOS field which is considered to have formed via a galaxy merger\citep{Scoville2007, Casey2023}. The imaging has revealed two compact nuclei with stellar masses of $\sim10^{11}$~M$_\odot$ separated by $\sim10$~kpc. Both nuclei have a prominent ring or shell around them, which has given rise to the name of the $\infty$ galaxy.  Chandra X-ray data and  Very Large Array radio data show the presence of an actively accreting black hole in the $\infty$ galaxy, with an X-ray luminosity of $L_X\sim1.5\times10^{44}$~erg/s and a rest-frame radio luminosity $L_{\rm 144 MHz}\sim2\times10^{26}$~W/Hz \citep{Dokkum2025a}. 

From the width of the H$\alpha$ line with $940\pm110$~km/s, together with the H$\alpha$ luminosity, they inferred a black hole mass of $\sim10^6$~M$_\odot$, which is roughly consistent with the observed fluxes assuming an Eddington accretion scenario. Interestingly, the position of the black hole does not correspond to one of the nuclei, but it appears to be at a position inbetween them. The active black hole is confirmed to be embedded in an ionized gas cloud and its radial velocity was shown to be very similar to the radial velocity in that cloud, making it difficult to reconcile the scenario with an ejection from the two nuclei \citep{Dokkum2025b}. Additionally, active black holes have also been identified in the two nuclei via H$\alpha$ and [N~II] emission, including the detection of broad lines, thus making it highly plausible that the massive black hole inbetween the nuclei should have a different origin.

The presence of gas as well as the efficient Eddington accretion make it highly plausible that gas has played a strong role during the formation. No star cluster is observed around the active black hole, though we cannot fully rule out that some remnants of a star clusters might be outshined by the radiation from the black hole. 

Now given the detection of very massive $\sim10^6$~M$_\odot$ and compact $\lesssim1$~pc clusters with JWST \citep{Adamo2024} that  consistent with the steep \citet{Marks2012} relation, it is a central and important question under what circumstances we could directly form a very massive black hole from the gas, or if instead a massive and compact cluster (such as the observed ones) should be set into place. We suggest here that perhaps the decisive distinction between the two cases could be related to the ratio between gas and stars during the evolution of the system, which might be regulated by the ratio between the gas inflow rate and the star formation rate. We expect the latter to be regulated by the interplay between gas inflow $\dot{M}$ and star formation. Assuming star formation to occur on the free-fall time $t_{ff}$ with an efficiency $\epsilon_{\rm SFR}$, we have\begin{equation}
\mathrm{SFR}=\epsilon_{\rm SFR}\frac{M_{\rm gas}}{t_{ff}}=\frac{\sqrt{8}}{\pi}\epsilon_{\rm SFR}\frac{\sqrt{G}M_g^{3/2}}{R^{3/2}},
\end{equation}
where we employed $t_{ff}=\sqrt{3\pi/(32G\rho_g)}$ and $\rho=3M_g/(4\pi R^3)$. The gas mass and stellar mass evolve as\begin{eqnarray}
\frac{dM_g}{dt}&=&\dot{M}-(1+\lambda)\mathrm{SFR},\\
\frac{dM_*}{dt}&=&\mathrm{SFR},
\end{eqnarray}
where $\lambda$ denotes the mass lost due to outflows related to star formation. The gas fraction $f_g=M_g/M_*$ thus evolves as\begin{equation}
\frac{df_g}{dt}=\frac{\dot{M}}{\mathrm{SFR}}-(1+\lambda).
\end{equation}
Requiring $\frac{df_g}{dt}>0$ then implies \begin{equation}
\dot{M}>\dot{M}_{\rm crit}={\rm SFR}(1+\lambda)=\frac{\sqrt{8}}{\pi}\epsilon_{\rm SFR}\frac{\sqrt{G}M_g^{3/2}}{R^{3/2}}(1+\lambda).\label{Mdotcrit}
\end{equation}
We assume here a cluster of size $R\sim$1~pc and a massive cloud with $M_g\sim10^6$~M$_\odot$. From Fig.~1 of \citet{Somerville2025}, we expect a star formation efficiency of $\epsilon_{\rm SFR}\sim1$ at these surface densities, implying a star formation rate of  $\sim 67\epsilon_{\rm SFR}$~M$_\odot$~yr$^{-1}$. Adopting $\lambda=1$ \citep[e.g.][]{Chisholm2017}, we thus have a critical inflow rate  $\dot{M}_{\rm crit}=134\epsilon_{\rm SFR}\,$M$_\odot$~yr$^{-1}$.

In the presence of such high accretion rates, the disk may be  unstable and prone to fragmentation \citep[e.g.][]{Lodato2007}. Nonetheless as a result of efficient gravitational torques, clumps are expected to migrate and merge with central objects on shorter timescales compared to KH timescale that is typically of the order of a few million years \citep{Escala2006, Inayoshi2014, LatifSchleicher2015}. Viscous heating may further stabilize the central part of the disk \citep{LatifSchleicher2015}. Additional torques can be expected in the presence of magnetic fields, which could further help to drive clumps into the center \citep{Latif2016d, Begelman2023,Latif2023, Latif2023b, Diaz2024}. We estimate the mass inflow rate due to gravitational torques as \citep{Lynden1972, Gammie2001, Escala2006, Escala2007, Hopkins2011, Inayoshi2014,LatifSchleicher2015}  \begin{equation}
\dot{M}_{G}\sim3\alpha_{\rm grav}\frac{v_{\rm rot}^3}{G},
\end{equation}
where we estimate the rotational velocity from the Kepler relation as $v_{\rm rot}=\sqrt{GM/R}$ and $\alpha_{\rm grav}\sim 0.03-0.3$, implying characteristic timescales $M/\dot{M}_G\sim1.6\times10^4-1.6\times10^5$~yr  for the system under consideration. Indeed the efficient operation of such gravitational torques was also confirmed via numerical simulations \citep[e.g.][]{Suazo2019}.

\begin{figure}[h!]
    \centering
    \includegraphics[width=0.8\hsize]{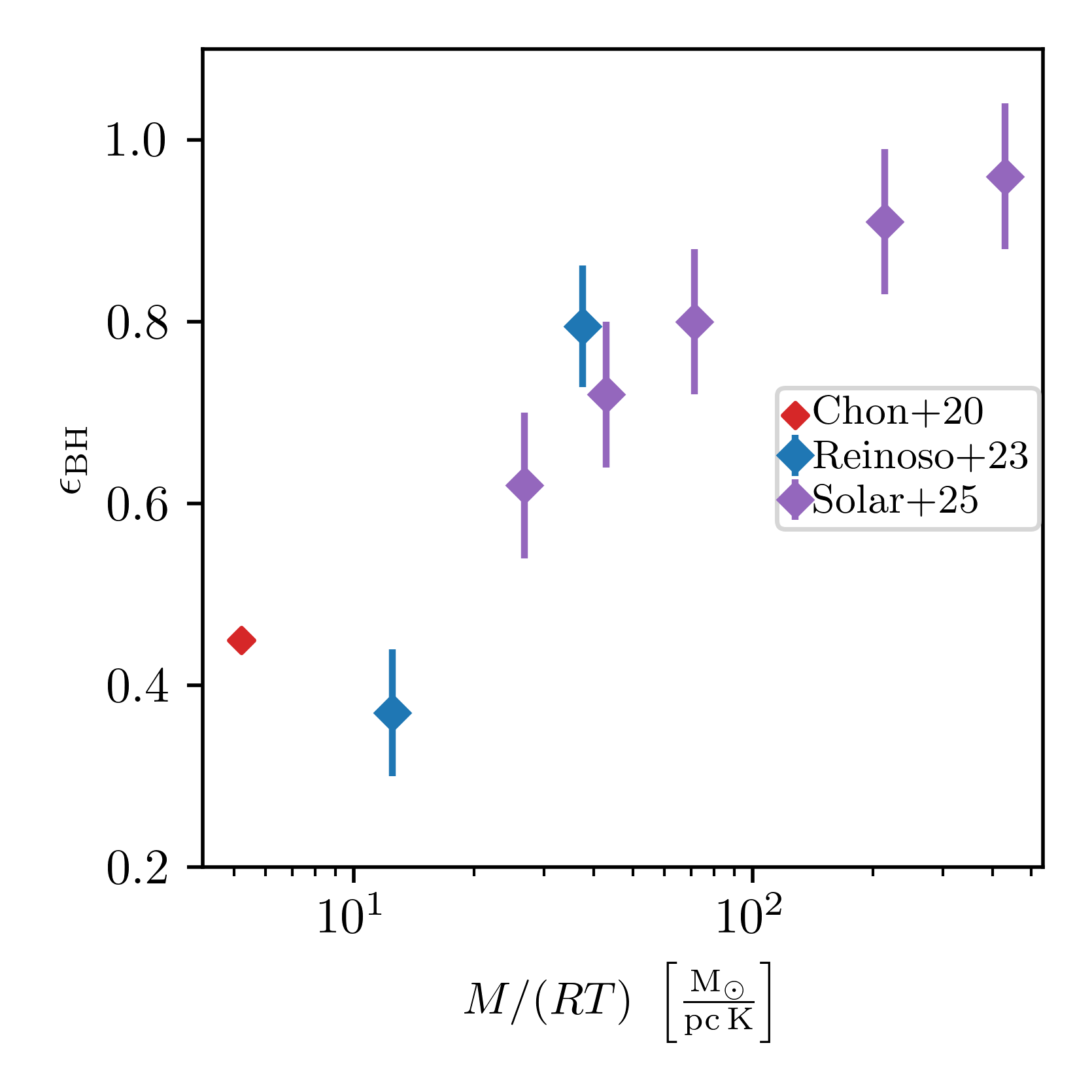}
\includegraphics[width=0.8\hsize]{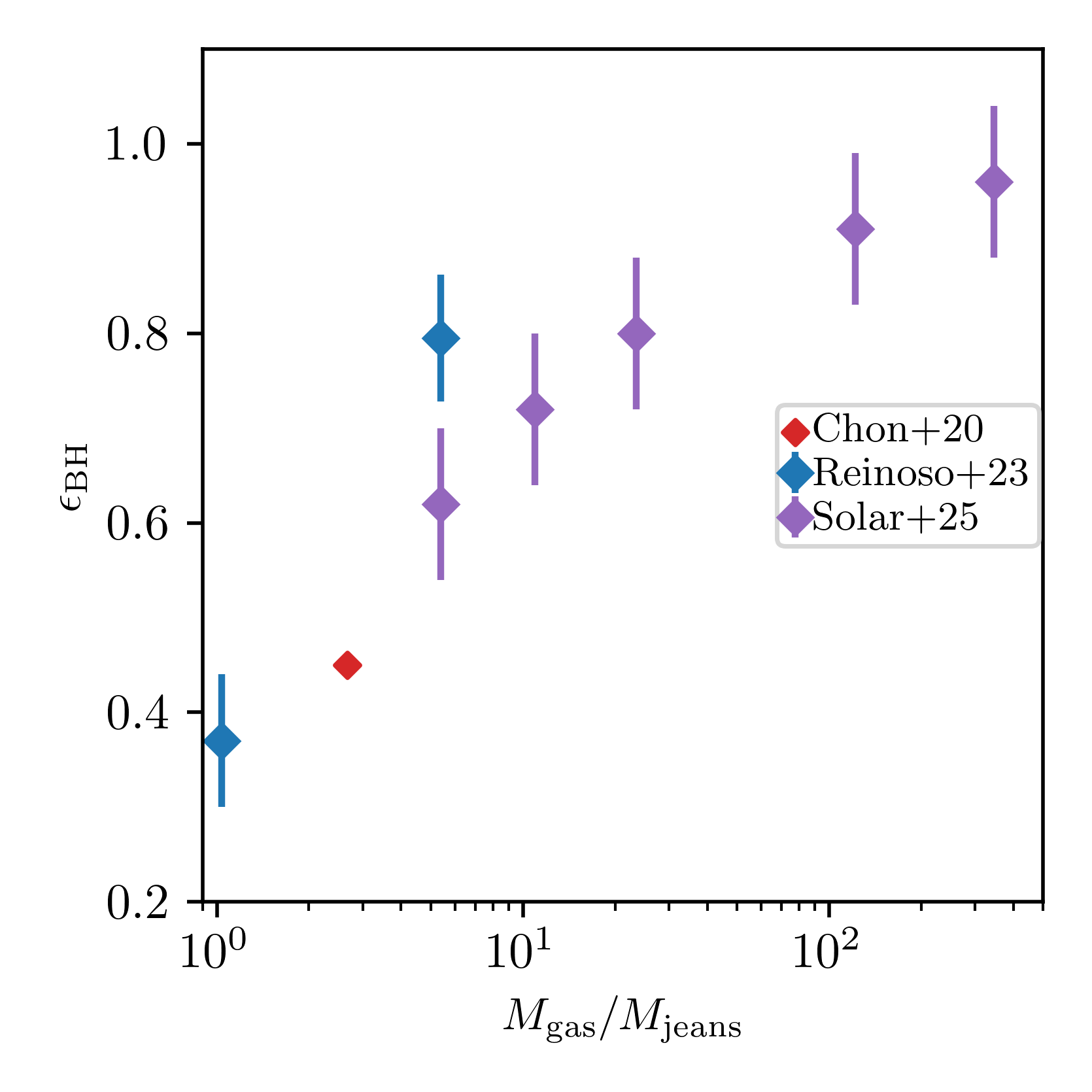}
    
    \caption{Top panel: Efficiency to form a central massive object (ratio of mass in that object over total mass) as a function of $M/(RT)$ to approximate the gravitational focusing parameter $\Phi$ as introduced in Eq.~\ref{phi}. Bottom panel: Here we show the efficiency to form a central massive object as a function of gas mass divided by thermal Jeans mass. In these plots we consider data points from simulations providing detailed models including collisions and accretion in massive gas-dominated clusters at sub-solar metallicities \citep{Chon2020, Reinoso2023, Solar2025}.}
\label{fig:Jeans}
\end{figure}

In a gas-dominated system, high accretion rates as mentioned above could be produced in at least a few different ways. A very low  metallicity $Z\lesssim10^{-3}$~Z$_\odot$ could be particularly favorable, as then the gas temperature $T$ is still high. From the scaling of the Jeans mass $M_J\propto T^{3/2}\rho^{-1/2}$ and the free-fall time $t_{ff}\propto\rho^{-1/2}$, one may then expect high accretion rates onto the star cluster as a result of the gravitational instability, with $M_J/t_{ff}\propto T^{3/2}$. \citet{Chon2020, Chon2025}  identified a regime where the mass growth is even dominated by a mode they refer to as super-competitive accretion \citep[see also][]{Schleicher2023}, while sometimes the contribution between mass growth and accretion can also be more balanced \citep[see][]{Solar2022, Reinoso2023, Solar2025}. Another possibility that could also operate at high metallicities is the occurrence of galaxy mergers, which can bring large amounts of gas to the center of the galaxy \citep{Mayer2015, Mayer2024}. Recent studies have shown that supermassive stars with accretion rates of $10-1000$~M$_\odot$~yr$^{-1}$, consistent with estimates from Eq.~\ref{Mdotcrit}, collapse into a massive black hole during hydrogen burning as a result of the general relativistic instability, leading to typical black hole masses of $\sim10^6$~M$_\odot$ \citep{Haemmerle2025}. A set of detailed simulations for the formation of central massive objects in gas-dominated systems has been compiled  by \citet{Solar2025} considering results from \citet{Chon2020, Solar2022} and \citet{Reinoso2023}. They found a strong relation between the efficiency to form a central massive object (defined as the ratio of its mass with respect to the total mass including gas mass) and the ratio between gas mass and thermal Jeans mass $M_J$. This relation is provided in Fig.~\ref{fig:Jeans}. As  $M_J\propto T^{3/2}\rho^{-1/2}$, we have \begin{equation}M/M_J\propto R^3 \rho^{3/2}T^{-3/2}\propto R^{-3/2}M^{3/2}T^{-3/2},\end{equation} with $\rho\propto MR^{-3}$. In case of a constant total mass and constant temperature, the relation suggests that compact systems will more efficiently form a central massive object.

\begin{figure}[h!]
   \centering
\includegraphics[width=\hsize]{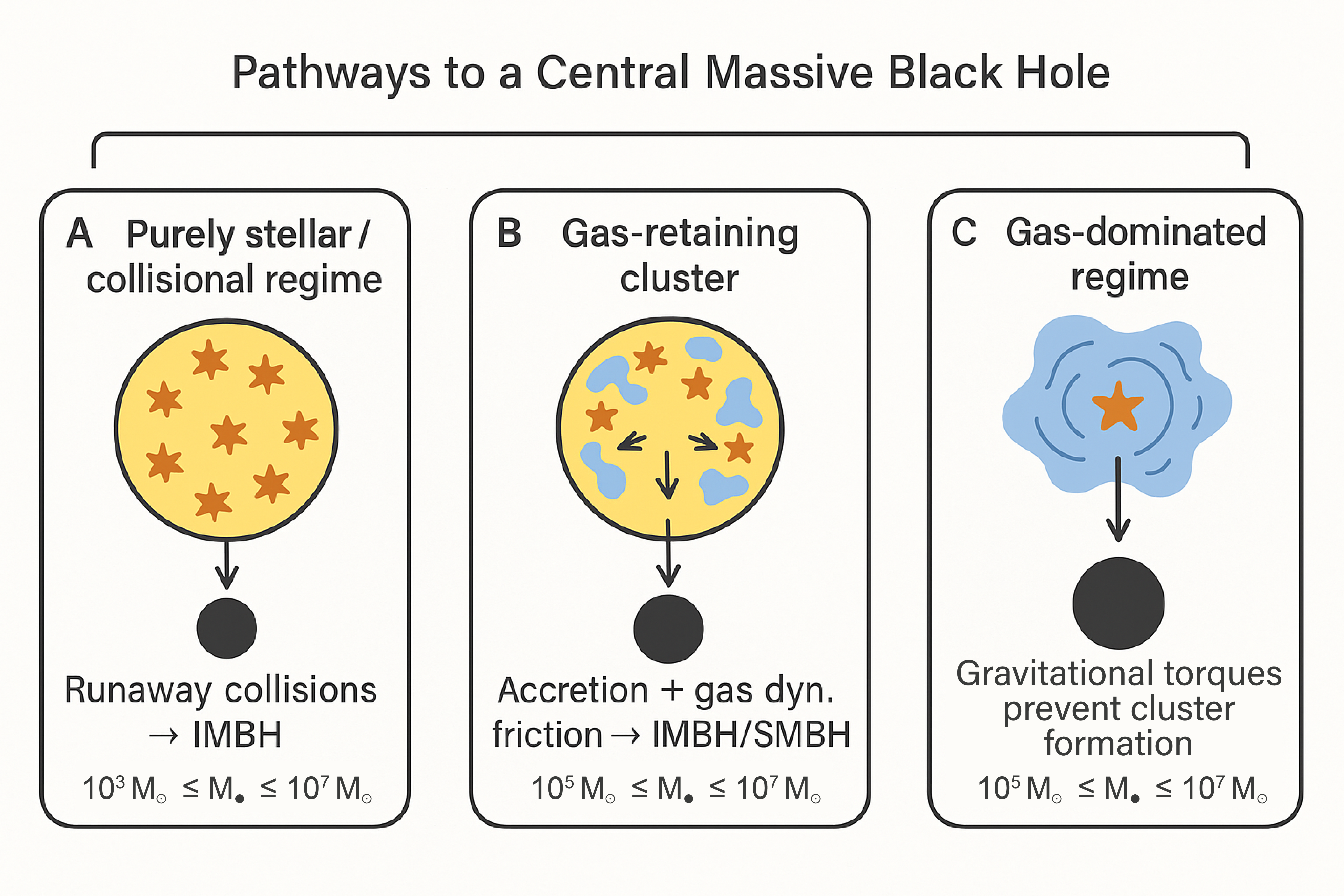}
 \caption{Summary of the black hole formation pathways discussed in this paper, describing the purely collisional regime (left), the case of gas retention in a massive compact cluster (mid panel) and the gas-dominated regime (right panel).}
         \label{channels}
   \end{figure}

Alternatively, this relation can be expressed via the gravitational focusing parameter\begin{equation}
\Phi=\frac{GM}{Rc_s^2},\label{phi}
\end{equation}
which describes the ratio between the depth of the potential well $GM/r$ and the thermal energy per unit mass $c_s^2\propto T$. We show in Fig.~\ref{fig:Jeans} that a very similar relation is found with this parameter as well. It is worth noting that the simulations we analyzed here correspond to simulations of sub-solar metallicity, studying the formation and evolution of massive clusters initially dominated by gas, and taking into account both the effects of accretion and collisions for the growth of central massive objects. Also \citet{Lahen2025} investigated the formation and evolution of star clusters in starburst dwarf galaxies at a metallicity of $\sim0.016$~Z$_\odot$, finding the formation of clusters with $2\times10^5$~M$_\odot$ and radii of $\sim1$~pc. Even though the stars in their model were only growing through collisions and accretion was not taken into account, still very massive stars of $\sim10^4$~M$_\odot$ formed in these clusters in spite of the higher metallicity. Very detailed simulations of star cluster formation and evolution at solar metallicity have been reported by \citet{Fujii2021}, considering cluster masses of $10^5$~M$_\odot$ but with large radii of $20$~pc. In these clusters the formation of a massive object has not been found, consistent with the criteria developed here. It thus  seems conceivable that such a gas-driven channel may work in a sufficiently gas-dominated environment. We provide a summary of the black hole formation channels considered in this paper, including the purely collision-dominated channel, gas retention in compact massive clusters and the gas-dominated channel in Fig.~\ref{channels}.

\section{Conclusions}\label{conclusions}

In this paper, we have systematically assessed the potential to form intermediate-mass black holes in the young massive star clusters (YMCs) discovered by JWST \citep{Vanzella2022, Vanzella2023, Adamo2024, Mowla2024}. As a starting point, we have assessed the formation mechanisms of massive clusters considering simulation results by \citet{Grudic2023} for a Milky Way galaxy, who were finding a mass-radius relation with a steep dependence on the mass but also a $1\sigma$ scatter of the cluster radii corresponding to roughly an order of magnitude and comparing to the even steeper relation by \citet{Marks2012} derived from a binary population analysis of well-known star clusters and a comparison with star clusters in the process of formation.  

We have compared both relations to the population of YMCs detected by JWST and to the young star clusters (YSCs) in the local Universe derived from the Legacy Extragalactic UV Survey (LEGUS) \citep{Brown2021}. We found that both types of star clusters seem to fit well the \citet{Grudic2023} relation including the expected scatter, providing a population of more compact clusters that are potentially well-suited to form intermediate-mass black holes. The \citet{Marks2012} relation provides a good fit particularly to the more compact systems among the YSCs and YMCs, thus providing useful constraints on the possible initial conditions of star clusters and their more compact configurations. We  discussed  the potential implications of the cooling physics at high redshift, where a lower metallicity could imply a higher characteristic density of fragmentation and therefore potentially even more compact clusters. 

Subsequently, we compared the populations of YMCs and YSCs with a sample of Milky Way Globular Clusters \citep{Baumgardt2018, Vergara2024}, providing a comparison with relevant timescales, including evaporation timescales, tidal disruption timescales, the relaxation time, the mass segregation time and both the global and the core collision timescale. We highlight here in particular that most of the YMCs detected by JWST have relaxation timescales considerably shorter than the age of the Universe and also still shorter than the age of the Universe corresponding to their redshift of detection. However, this sample of clusters includes also some very massive systems ($10^7$~M$_\odot$) with cluster radii of up to $20$~pc, implying that some of them would not be able to relax even over the full age of the Universe. A relevant fraction of the YMCs further have global and core collision timescales of the order of $14$~Gyrs, which of course will decrease significantly once the system will go through core collapse.

A still relevant fraction of the population shows compact radii of $\sim0.7$~pc close to the \citet{Marks2012} relation. These have very short relaxation times and even their collision timescales are comparable to the age of the Universe. In case of a core collapse, it is strongly implied that efficient runaway-collisions will happen, as described e.g. by \citet{Poregies2002, Reinoso2018, Sakurai2019, Escala2021, Vergara2023, Vergara2025b}.

We have subsequently assessed the expected formation timescales for black hole seeds in the collision-based channel as a function of the expected size of the cluster. Assuming sufficiently dense clusters for an efficiency of order $10\%$, the cluster mass should be about $10$ times the black hole mass. Small seeds of $10^3$~M$_\odot$ can then be formed efficiently in clusters with typical sizes of up to $0.1$~pc. In case of $10^5$~M$_\odot$ seed masses, typical cluster sizes of $0.2$~pc are still sufficient, while in case of very massive seeds of $10^7$~M$_\odot$, the formation could occur in clusters with radii of up to $1$~pc. Comparing to the \citet{Grudic2023} mass-radius relation, these limits correspond roughly to the edge of the $1\sigma$ deviation, suggesting that about $16\%$ of real star clusters could be expected in this regime.

For particularly massive clusters following the \citet{Marks2012}, as some of the systems observed by JWST \citep{Adamo2024}, we derived a characteristic mass scale of $\sim6\times10^6$~M$_\odot$ above which supernova feedback is expected to become inefficient and the gas should thus be retained in such systems. Based on an analysis of the gas dynamical friction time and the Bondi accretion time in the core, we have shown that a massive black hole of $10^5-10^6$~M$_\odot$ should be able to efficiently form in such systems.

We finally considered the implications of a formation channel even more strongly dominated by gas, in cases where the gas inflow rate dominates over the star formation rate, therefore providing a continuously increasing gas to stellar mass ratio. We have derived a critical inflow rate of $\sim137\epsilon_{\rm SFR}$~M$_\odot$~yr$^{-1}$, with $\epsilon_{\rm SFR}$ the star formation efficiency, and shown that in such a case gravitational torques are expected to prevent the formation of a cluster and instead favour the formation of a central massive object. Such high inflow rates could be expected particularly in cases of low metallicity where the gas temperature is high \citep[e.g.][]{Omukai2005}, or in cases of galaxy mergers where large amounts of gas my flow to the central region of the galaxy \citep{Mayer2015, Mayer2024}. Such a gas-dominated channel has been investigated via numerical simulations by \citet{Chon2020, Solar2022, Reinoso2023} and \citet{Solar2025}, and the results suggest that the efficiency to form central massive objects in such gas-dominated clusters is strongly related to the ratio of gas mass over Jeans mass, or alternatively could be expressed in relation to the gravitational focusing parameter, which scales proportional to cluster mass over cluster radius and temperature for the systems considered here. In compact systems this channel can reach efficiencies in the range of $30-90\%$. Work by \citet{Haemmerle2025} has shown that supermassive stars with accretion rates of $10-1000$~M$_\odot$~yr$^{-1}$, consistent with the critical inflow rate mentioned above, indeed collapse into a massive black hole during hydrogen burning as a result of the general relativistic instability. 

While in the general case it is hard to determine whether a supermassive black hole has originated from a gas-based or a collision-based channel, we here considered the situation of the $\infty$ galaxy reported by \citet{Dokkum2025a, Dokkum2025b}, where an active black hole with $\sim10^6$~M$_\odot$ has been detected in a galaxy that also hosts two stellar nuclei of about $10^{11}$~M$_\odot$. Based on kinematics, the $\sim10^6$~M$_\odot$ appears incompatible with the idea of originating from these nuclei, which were found to host their own intermediate-mass black holes \citep{Dokkum2025b}. Instead the $\sim10^6$~M$_\odot$ appears to be embedded in an ionized gas cloud and seems to accrete roughly at Eddington luminosity, suggesting that gas was likely already present at its formation and may have even dominated its formation channel. This system may provide a relevant example for a massive black hole to have formed from a gas-based channel where gravitational torques induced by strong gas inflow have prevented the formation of even a compact star cluster.

Overall, we conclude that at least a fair fraction of the YMCs detected by JWST are likely to form intermediate-mass black holes with masses of $\sim10^5$~M$_\odot$ or potentially higher. This mass range has been suggested by numerical simulations from two different groups, \citet{Vergara2025b} and \citet{Rantala2025}. The formation of such massive seeds is highly favorable to explain some of the most massive black holes detected at high redshift \citep{Sassano2021, Trinca2022, Spinoso2023} and can also lead to important gravitational wave signatures for future gravitational wave detectors such as the Laser Interferometer Space Antenna (LISA)\footnote{LISA: https://lisa.nasa.gov/} \citep{Valiante2021}.

\begin{acknowledgements}
{We thank the anonymous referee for very constructive and valuable comments on our manuscript.}    DRGS  gratefully acknowledges support from the Alexander von Humboldt - Foundation, Bonn, Germany.  ML gratefully acknowledges support from ANID/DOCTORADO BECAS CHILE 72240058.   MCV acknowledges funding through ANID (Doctorado acuerdo bilateral DAAD/62210038) and DAAD (funding program number 57600326). DRGS, AE, JSB and FC thank for funding via the ANID BASAL project FB21003. PS acknowledges support through ANID/Doctorado en el Extranjero convocatoria 2022 (funding number 72220198) and thanks the German Federal Ministry of Research, Technology and Space and the German federal states (http://www.nhr-verein.de/en/our-partners) for supporting this work as part of the National High-Performance Computing (NHR) joint funding program, and the Kultrun cluster hosted at the Departamento de Astronomía, Universidad de Concepción. For this work the HPC-cluster Hummel-2 at University of Hamburg was used. The cluster was funded by Deutsche Forschungsgemeinschaft (DFG, German Research Foundation) – 498394658.
MG was supported by the Polish National Science Center (NCN) through the grant 2021/41/B/ST9/01191.
This material is based upon work supported by Tamkeen under the NYU Abu Dhabi Research Institute grant CASS. 
 BR acknowledges support by the European Research Council via ERC Consolidator grant KETJU (no. 818930)
AA acknowledges support for this research from the Polish National Science Center (NCN) grant number 2024/55/D/ST9/02585. For the purpose of Open Access, the authors have applied for a CC-BY public copyright license to any author Accepted Manuscript (AAM) version arising from this submission.
JSB acknowledges funding through ANID (Doctorado acuerdo bilateral DAAD/62240030) and DAAD (funding program number 57752768).
\end{acknowledgements}

\end{document}